# A Miniature Non-Uniform Conformal Antenna Array Using Fast Synthesis for Wide-Scan UAV Application

Yuanyan Su, Icaro V. Soares, *Student Member, IEEE*, Siegfred Daquioag Balon, Jun Cao, Denys Nikolayev, and Anja K. Skrivervik

*Abstract*—To overcome the limited payload of lightweight vehicles such as unmanned aerial vehicle (UAV) and the aerodynamic constraints on the onboard radar, a compact non-uniform conformal array is proposed in order to achieve a wide beamscanning range and to reduce the sidelobes of the planar array. The non-uniform array consists of 7×4 elements where the inner two rows follow a geometric sequence while the outer two rows follow an arithmetic sequence along the *x*- axis. The element spacing along the *y*-axis is gradient from the center as well. This geometry not only provides more degrees of freedom to optimize the array radiation, but also reduces the computation cost when synthesizing the excitation and the configuration of the array for a specific beam pattern. As field cancellation may happen due to the convex and concave features of the non-canonical UAV surface, a fast and low-cost in-house code to calculate the radiation pattern of a large-scale conformal array for an arbitrary surface and element pattern is employed to optimize the array structure. As a proof of concept, the proposed array with a total volume of 142×93×40 mm$^3$ is implemented at ISM band (~5.8 GHz) using a miniature widebeam single-layer patch antenna with a dimension of 0.12$\lambda_0$×0.12$\lambda_0$×0.025$\lambda_0$. By using the beamforming technique, an active onboard system is measured, which achieves the maximum gain of 21.8 dBi and a scanning range of >50° and –28°~28° with a small scan loss of 2.2 and 0.5 dB in elevation and azimuth, respectively. Therefore, our design has high potential for wireless communication and sensing on UAV.

## I. Introduction

FOR modern remote sensing radars and airborne communication which require a small aircraft or unmanned aerial vehicle (UAV), conformal array antennas are more desirable for aerodynamic purposes. Compared to planar arrays, conformal arrays can be easily accommodated on a non-flat platform such as the wing, nose, base, and tail of the aircraft, with no need for additional space. In the past decades, most of the developed conformal antenna arrays have the convex shapes, for example, cylindrical arrays [1-13], spherical arrays [13-17], hemispherical arrays [18-20] and wing-shaped arrays [21-23]. The main advantages of these convex conformal arrays are to reduce the blind zone of planar arrays and to widen the scanning angle when the array element has directive radiation, since a convex surface allows more elements to take over the directivity at the scanning edge. Also, for an electrically large array with hundreds of elements, this type of conformal array could contain more elements than the planar array having the same projection area, thus providing more opportunities to improve the antenna performance.

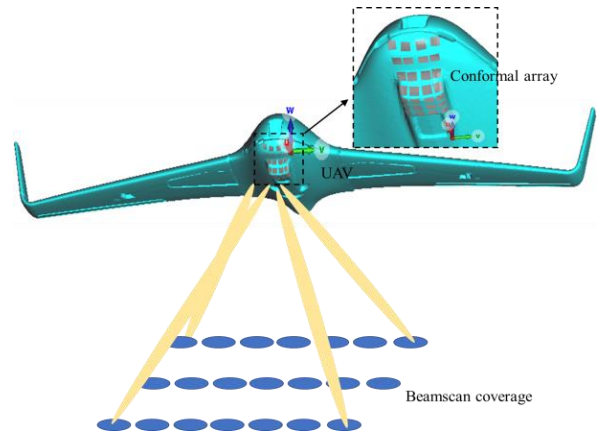

Fig. 1. Illustration of a conformal antenna array mounted on a UAV with the convex and concave features.

However, in some realistic circumstances, a non-canonical surface might be used more often due to the structural design of the UAV itself. For instance, the antennas need to be installed beneath the front head of the aircraft, as shown in Fig. 1. Such a general but sophisticated surface has both convex and concave features, resulting in a more complicated environment for the array element to radiate effectively. Besides, array factor is no longer valid in the conformal array synthesis, a high mutual coupling may deteriorate the antenna impedance, and field cancellation would require the array system to be reconfigured. To the best of the authors' knowledge, there is little discussion on designing, analyzing, and evaluating this type of conformal antenna array in the literature. A preliminary study based on [24] was conducted for the non-canonical surface, showing it as an example to demonstrate the feasibility of the calculation method. Another design using discrete circular patch elements conformed on a concave hemispherical-like surface is presented in [25], where the extra mechanical support and fixture are required to install the flat array elements on the curved surface.

Besides the geometry constraint on the non-canonical surface, the selection of the array element is also critical to a compact conformal array design which requires a broad-beam and high-gain radiation element to achieve a wide scanning coverage. Unlike the variety of options available for planar arrays, conformal arrays are now mostly using dipoles and patch antennas, or their variants, since the curvature of the conformed surface would increase the mechanical design difficulty of the array element and weaken the system's robustness. Moreover, there is a trade-off between array miniaturization and excellent radiation performance for a space-limited UAV. The decrease in antenna gain and working bandwidth and the increase in the sidelobe level (SLL) should be particularly taken care of when miniaturizing the array element. In [26], the authors proposed a simple single-layer linear-polarized microstrip antenna, the size of which is around 0.23% wavelength only. Ref. [27] shows a well-behaved miniature antenna that is able to lower the operating frequency by introducing more capacitance with the etched slots. A simpler design based on the multimode principle is provided in [28], but it seems difficult to combine these modes on a smaller aperture. In general, there are plenty of miniature antennas investigated in the past [29-37]. However, very few of them are solved by utilizing an extremely simple and small structure, in the order of ~$0.1\lambda_0$, for a conformal array on UAV.

In this paper, a compact non-uniform array conformed to a non-canonical UAV surface is first proposed and implemented at 5.8 GHz to obtain a wide two-dimensional (2D) beamscanning coverage with reduced SLLs. The improved, fast and efficient in-house code to calculate the radiation pattern of an arbitrarily-configured array with the pre-defined element pattern is particularly helpful to synthesize such array and even a large-scale array for different applications. In addition, to explore the potential performance improvement, a convex optimization method based on the developed code is used to obtain the optimal amplitude weighting coefficients of the array elements. It is shown that weighting the amplitude of individual elements indeed would further reduce the SLL, but drop the antenna gain, widen the beamwidth, and increase the complexity of the beamforming feeding network (BFN). Therefore, as a proof of concept, the prototype of the non-uniform conformal array with an equal input power is first experimentally validated.

This paper is organized as follows. The proposed conformal array synthesis method, as well as a review on the algorithm to realize the fast radiation pattern calculation, is presented in Section II. In this section, the proposed array is also compared with other common arrays in theory. The design and simulation of the array element and the conformal array are then described in Section III. This section also presents the designs of the feeding network and the whole onboard system with the aid of a full-wave simulation. The details of the measurements and the results of the fabricated array system are discussed in Section IV. The conclusion is finally drawn in Section V.

## II. DESIGN THEORY

Many synthesis methods have been reported in designing a conformal array, such as projection-based methods [38, 39], iterative least-squares [40], convex optimization [41, 42], genetic algorithm [43-45], particle swarm [46], and hybrid methods [38, 47-48]. However, these methods are unclear when applied to a non-conventional array for the purpose of achieving a wide 2D beamscanning coverage and reduced SLL simultaneously. Therefore, we propose a parametric non-uniform conformal array that is synthesized using an improved code to accelerate the calculation of the 3D radiation. The mathematics is presented in the beginning of this section. With this knowledge, the non-uniform conformal array is designed theoretically. Its performance is finally compared to a conformal array with uniform spacing, and a planar array with uniform spacing to demonstrate the merits of the design.

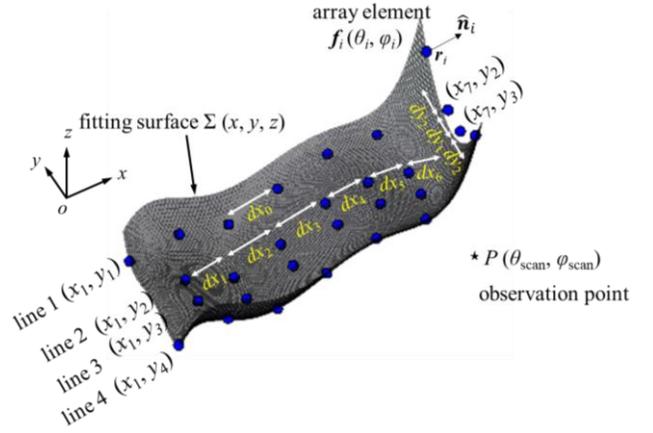

Fig. 2. Definition and modeling of the non-canonical UAV surface and formulation of the synthesis problem.

### A. Algorithm to Calculate Radiation Pattern

The algorithm to calculate an array pattern starts from the code developed in [24]. First, a sophisticated UAV surface, as illustrated in Fig. 1, is mathematically analyzed by integrating a polynomial fitting method with the extraction of element position and normal vector on the surface. The polarization of the array element can be assigned, which allows for the arbitrary radiation pattern of a polarized element to be included in the calculation. For example, a patch antenna with linear polarization and $\cos^2$ pattern that will be realized in Section III, is included as one of the element types in this improved algorithm.

Since the position $r_i$ and normal vector $\hat{n}_i$ of each element are the prerequisites of the calculation, polynomial fitting is the best to model such concave and convex surface. Different from the interpolant fitting functions (linear, spline, B-spline, etc.) that may give a better approximation but lack an explicit expression, it can provide an accurate analytical formula for $r_i$ globally when the sampling points on the surface are sufficient. The geometry of the fitting surface is shown in Fig. 2, where the ranges in $x$ and $y$ directions are set as 96~221 mm and -37.6~37.6 mm with a mesh grid step of 0.5 mm, respectively. A small root mean square error (RMSE) of 0.002 is achieved from the polynomial fitting result, hence evidences an accurate approximation to the actual surface. The polynomial fitting function is expressed as

$$z = f(x,y) = p_{00} + p_{10}x + p_{01}y + p_{20}x^2 + p_{11}xy + p_{02}y^2$$
$$+ p_{30}x^3 + p_{21}x^2y + p_{12}xy^2 + p_{03}y^3$$
$$+ p_{40}x^4 + p_{31}x^3y + p_{22}x^2y^2 + p_{13}xy^3$$
$$+ p_{04}y^4 + p_{50}x^5 + p_{41}x^4y + p_{32}x^3y^2$$
$$+ p_{23}x^2y^3 + p_{14}xy^4 + p_{05}y^5 \quad (1)$$

where the polynomial coefficients are summarized in Table I. The blue particles in Fig. 2 indicate the array elements characterized by the normal vector $\hat{\boldsymbol{n}}_i = (\theta_i, \varphi_i)$ and the element position $(x_i, y_i)$ that is related to the element number and element spacings $dx_i$ and $dy_i$.

TABLE I
POLYNOMIAL COEFFICIENTS OF THE FITTING FUNCTION FOR THE NON-CANONICAL UAV SURFACE

| Coeff. | Value | Coeff. | Value |
|---|---|---|---|
| $p_{00}$ | −1.55797797802342 | $p_{20}$ | −619.894754098413 |
| $p_{01}$ | 2.91329870656138e−05 | $p_{21}$ | 0.00727428458918239 |
| $p_{02}$ | −2.82550052149489 | $p_{22}$ | 4228.10444461165 |
| $p_{03}$ | −0.000137746419535819 | $p_{23}$ | 0.00255991407797952 |
| $p_{04}$ | 2704.54313735367 | $p_{30}$ | 3816.19583103342 |
| $p_{05}$ | 0.00341547370121415 | $p_{31}$ | −0.0303751734433911 |
| $p_{10}$ | 48.8878266621642 | $p_{32}$ | −14489.3530756313 |
| $p_{11}$ | −0.000759694335898899 | $p_{40}$ | −11480.5566776082 |
| $p_{12}$ | −176.329237389226 | $p_{41}$ | 0.0466945380798176 |
| $p_{13}$ | 0.000807535950634196 | $p_{50}$ | 13628.3459462032 |
| $p_{14}$ | −32267.8360780080 | | |

As each element of this conformal array sees a different environment, the array factor is not applicable to its farfield pattern calculation. Instead, the pattern can be derived from the superposition of the contribution of each individual element:

$$\boldsymbol{E}(\theta,\varphi) = \sum_{i=1}^{N\times M} \boldsymbol{f}_i w_i e^{-jk\boldsymbol{r}_i\cdot\boldsymbol{r}_p} \quad (2)$$

where $\theta$ is the azimuth angle;
 $\varphi$ is the elevation angle;
 $f_i$ is the radiation pattern of the individual element;
 $k = 2\pi/\lambda$ is the wavenumber at frequency $f$;
 $\boldsymbol{r}_i = (x_i, y_i, z_i)$ is the position vector of the $i^{th}$ element relative to the origin of the global coordinate system (GCS);
 $\boldsymbol{r}_p$ is the vector of an arbitrary observation point $P(\theta, \varphi)$ relative to the origin of GCS.

For the isotropic elements whose $f_i$ is a scalar constant independent of $(\theta, \varphi)$, the GCS can be directly used. However, for an arbitrary conformal array consisting of directive elements, it is more convenient to use the local coordinate system (LCS) and normal vector $\hat{\boldsymbol{n}}_i$. By using the coordinate transformation matrix $\mathbf{T}$ as in [24], the set of observation point vectors $\{(\boldsymbol{r}_{p1},\ldots,\boldsymbol{r}_{pi})| \boldsymbol{r}_{pi} \in \mathbf{R}^3\}$ for each element can be derived as

$$\boldsymbol{r}_{pi} = \sum_{m=1}^{3} \mathbf{R}_m, \quad \mathbf{R}_m = \mathbf{T}\boldsymbol{r}_p \quad (3)$$

To accomplish the beam scanning, the complex weights of $w_i = a_i e^{-jk\boldsymbol{r}_i\cdot\boldsymbol{r}_s}$ where $a_i$ is the input power amplitude of the $i^{th}$ element and $\boldsymbol{r}_s$ is the beam steering vector defined in (4), are required to compensate the path difference for the equal phase front in the far field.

$$\boldsymbol{r}_s = (-\cos(\varphi_{scan})\cos(\theta_{scan}), -\cos(\varphi_{scan})\sin(\theta_{scan}), -\sin(\varphi_{scan})) \quad (4)$$

The improvements of the algorithm in [24] are done and two modifications should be highlighted. One is the inclusion of the amplitude compensation for the directive elements, since the definition of steering vector in (4) is only true for the isotropic elements. The other modification is concerning the element type. The options include not only the ideal isotropic source, but now also the polarized/non-polarized dipole with an isotropic pattern in H-plane, the polarized/non-polarized $\cos^2$ pattern, and the other directive antennas with user-defined polarization. The principle to consider the antenna polarization is basically to project the field components in two orthogonal planes, then to calculate the farfield radiation independently, and finally to sum up these two components together. Considering an arbitrary conformal array that comprises the elements with a radiation pattern $\boldsymbol{f}_i'$:

$$\boldsymbol{f}_i' = f_i \hat{\boldsymbol{p}}_i' \quad (5)$$

where $\hat{\boldsymbol{p}}_i'$ is the polarization unit vector defined in LCS, the projection of each polarization vector in GCS can be found by taking the inverse matrix of the coordinate transformation $\mathbf{T}^{-1}$, which is written as:

$$\boldsymbol{f}_i = \mathbf{T}_i^{-1} \cdot (f_i \hat{\boldsymbol{p}}_i'). \quad (6)$$

Therefore, the total field can be calculated using the superposition in (2), and the directive properties of the array can be obtained as well.

A directive pattern with a mathematical form of $f_i = \cos^2\varphi_i\cos^2\theta_i$ is desired and used as a case study in this section. This is because, on the one hand, it can provide a higher antenna gain even for a small-aperture array. On the other hand, the ground plane of the patch-type antenna can shield the large UAV background to a certain extent, thus could alleviate the negative effects such as gain reduction, pattern distortion, and the increases in SLL and antenna coupling. To summarize, the flowchart of the improved algorithm to calculate the radiation performance of an arbitrary conformal array with the user-defined element pattern is drawn in Fig. 3.

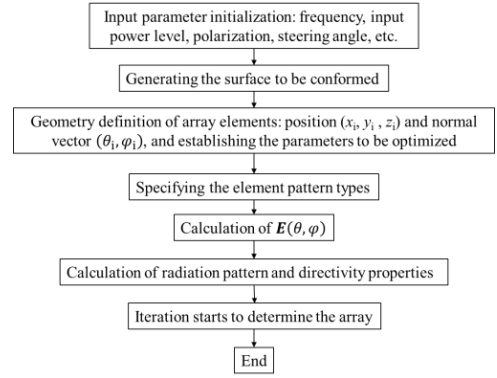

Fig. 3. Flowchart of the improved algorithm to calculate the radiation performance of an arbitrary conformal array with a user-defined element.

TABLE II
OPTIMIZED GEOMETRY OF THE NON-UNIFORM CONFORMAL ARRAY

| Parameter | Value | Parameter | Value |
|---|---|---|---|
| $dx_1$ | 20.7 mm | $dy_1$ | 27 mm |
| $x_1$ | 107 mm for Lines 2 and 3<br>96.1 mm for Lines 1 and 4 | $dy_2$ | 24.1 mm |
| $q$ | 0.98 | $N_{y\times x}$ | 4×7 |

Furthermore, based on this algorithm, we intend to investigate a global optimization of the amplitude weights by the use of a convex optimization method to achieve a broad 2D beamscanning coverage and reduced SLL, such that only the phase is the variable to be controlled through phase shifters. The weighting coefficient to be optimized for each element is defined as

$$\mathbf{w} = [w_1, w_2, \ldots, w_N]. \quad (7)$$

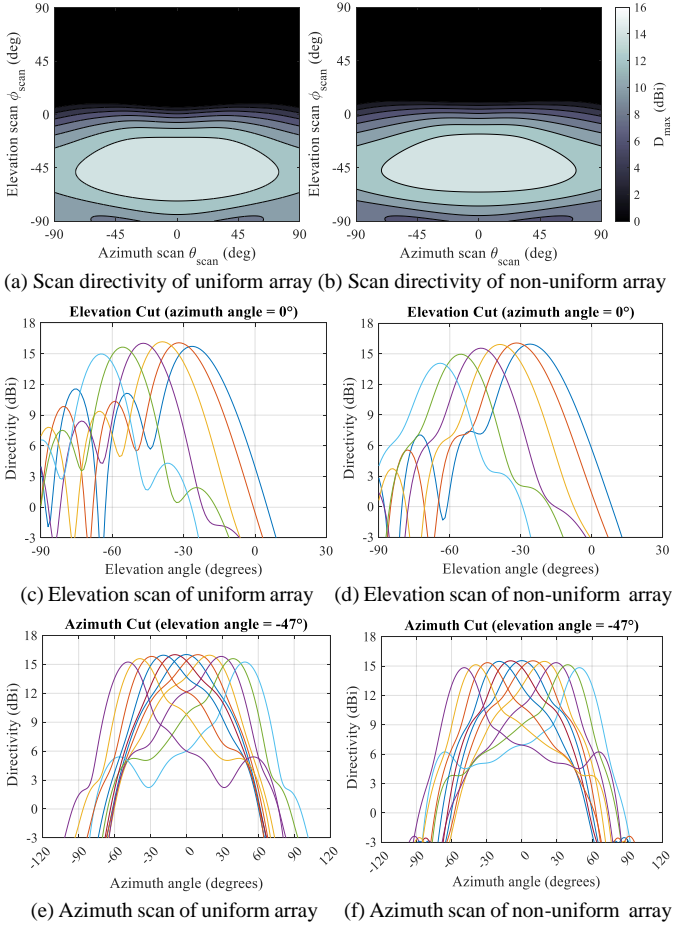

Fig. 4. Comparison between the non-uniform conformal array and the uniform conformal array in terms of the 2D scanning directivity contour, elevation scan, and azimuth scan at 5.8 GHz.

According to the directivity definition in (8),

$$D = \frac{4\pi U(\theta,\varphi)}{P_{rad}} \quad (8)$$

and assuming that the change of the total radiation power is negligible, the only variable related to $\mathbf{w}$ is the radiation intensity $U(\theta, \varphi)$ which has the form of

$$U(\theta, \varphi) = |\mathbf{w} \cdot \mathbf{E}(\theta, \varphi)|^2. \quad (9)$$

The electric field $\mathbf{E}(\theta, \varphi)$ in (9) is using the same definition in (2), except that the complex weights used to compensate for the steering angle in (2) become $e^{-jk\mathbf{r}_i \cdot \mathbf{r}_s}$ instead, as they are assumed to have the equivalent unity amplitude before the optimization. Thus, $\mathbf{E}(\theta, \varphi)$ can be re-written as

$$\mathbf{E}(\theta, \varphi) = \sum_{i=1}^{N \times M} \mathbf{f}_i \, e^{-jk\mathbf{r}_i \cdot \mathbf{r}_s} e^{-jk\mathbf{r}_i \cdot \mathbf{r}_p}. \quad (10)$$

In other words, the sum of the weighted field from $N$ antenna elements can represent the response of the antenna array at a certain observation point. SLL is an important index to evaluate the radiation performance of the array, and can be defined as the ratio of the radiation level in the sidelobe region $U(\theta, \varphi)$ (where $(\theta, \varphi) \in$ a defined sidelobe region) to the maximum radiation level in the mainlobe region $U_{max}(\theta_{scan}, \varphi_{scan})$. Then, the goal is set to minimize the SLL and maximize the radiation level for each steering angle. The optimization problem with the constraints is mathematically described as below.

*min* SLL

*S.t.*

$\frac{|\mathbf{w} \cdot \mathbf{E}(\theta,\varphi)|^2}{U_{max}(\theta_{scan},\varphi_{scan})} \leq$ SLL, $(\theta, \varphi) \in$ defined sidelobe region (11)

$|\mathbf{w} \cdot \mathbf{E}(\theta,\varphi)|^2$ monotonically decreases from $|\mathbf{w} \cdot \mathbf{E}(\theta_{scan}, \varphi_{scan})|^2$, where $(\theta, \varphi) \in$ mainlobe region (12)

$|\mathbf{w} \cdot \mathbf{E}(\theta_{scan}, \varphi_{scan})|^2 \geq 0.707 \cdot U_{max}^{ref}$ (13)

SLL $\leq 0.1$ (14)

### B. Synthesis of Non-Uniform Conformal Array

Since the available antenna system space is only 130×80 mm² in *xy*-plane and the element spacing must be larger than the element size, the number of array elements, particularly in *y* direction, is difficult to exceed the maximum of 4, thus weakening the scanning ability and reducing the freedom to optimize the radiation pattern on that plane. According to our study (see Section II-C), the uniform conformal array having equal spacings along *x* and *y* directions would suffer from a high SLL of ~ -4.1 dB at some observation angles. This high SLL would affect the radiation performance on the lower steering angles and should be reduced. In contrast, this sidelobe suppression can be accomplished using a non-uniform conformal array, since it increases the design degree of freedom to control the farfield radiation.

The geometry of the proposed conformal array with a non-uniform distribution is depicted in Fig. 2. As can be seen, the element spacing along *x* and *y* directions is designed independently. In the *x*-direction, the position of the first elements in the two inner rows, i.e., ($x_1$, $y_2$) and ($x_1$, $y_3$), needs to be defined, and the element spacing follows a geometric sequence as:

$$dx_n = dx_1 \cdot q^{n-1}, (n = 2, \ldots, N-1), \quad (15)$$

in which the initial term $dx_1$ is the first element spacing, $q$ is the common ratio, and $N$ is the element number. Then, the *x*-axis coordinates of the rest of elements in the two inner rows are derived using:

$$x_{n+1} = x_n + dx_n, (n = 1, \ldots, N-1). \quad (16)$$

In contrast, the location of the elements on the outer rows obeys an arithmetic sequence with a common difference $dx_0$ equal to

$$dx_0 = \frac{\max(x) - \min(x)}{N-1}, \quad (17)$$

where $\mathbf{x}$ is the *x*-axis vector of the fitting surface, which means that the first and end elements on the edges must locate on the boundary of the fitting surface.

In the *y*-direction, the spacing between the inner two-row elements is denoted as $dy_1$, different from the spacing $dy_2$ between the inner elements and their neighboring edge elements. In this way, the array symmetry along *y*-direction is still guaranteed and a fully parametric design to the non-uniform conformal array is established. Finally, these parameters, as tabulated in Table II, can be optimized through the iterations with the developed algorithm in Section II-A.

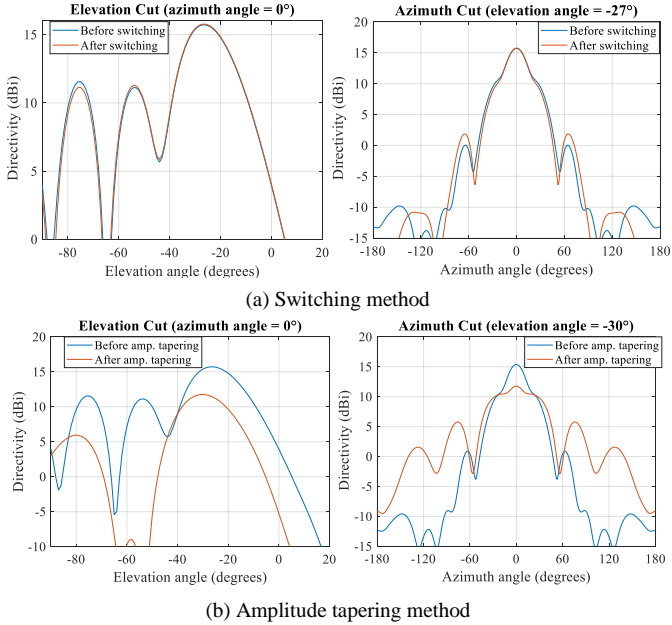

(a) Switching method

(b) Amplitude tapering method

Fig. 5. Pattern comparison of the uniform conformal array before and after applying the switching method and the amplitude tapering method at the steering angle of ($\theta_{scan} = 0°$, $\varphi_{scan} = -27°/-30°$).

## C. Non-Uniform Conformal Array vs. Uniform Conformal Array

To demonstrate the advantages of the proposed non-uniform conformal array, a uniform conformal array with equal spacings of $dx$ = 20.8 mm and $dy$ = 25 mm along $x$ and $y$ directions is discussed for comparison. Fig. 4 plots the 2D scanning directivity contour and compares the scanning patterns of these two conformal arrays in two cutting planes. It is found that the uniform conformal array achieves the azimuth scan range of $-49° \leq \theta_{scan} \leq 49°$ and the elevation scan range of $-65° \leq \varphi_{scan} \leq -26°$ with a small scanning loss of < 0.8 dB and < 1.2 dB, correspondingly. The maximum directivity occurs at the steering angle of ($\theta_{scan} = 0°$, $\varphi_{scan} = -43°$). However, its SLL in elevation scan is particularly high, reaching up to $-4.1$ dB within the scanning range. In contrast, the non-uniform array can reduce the highest SLL by 4.4 dB in the elevation scan and obtain the same scanning coverage with a comparable scanning loss. Unlike the uniform array, its highest sidelobe in the azimuth scan almost moves out of the required scanning range, thus less affecting the scanning radiation performance. Besides, the proposed non-uniform array has a high potential to shrink the large sidelobe tail that happens in the high steering angles of the uniform array.

In order to mitigate the high SLL of the uniform array, the switching method, which is to switch off the elements with a trivial radiation contribution, and the amplitude tapering method that is to turn off the less-effect elements and re-distribute the input power equally, are also investigated to configure the excitation of the uniform array. The corresponding farfield patterns in the steering angle of ($\theta_{scan} = 0°$, $\varphi_{scan} = -27°/-30°$) are presented in Fig. 5. As can be seen, no matter which method is applied to the uniform array, the SLL decreases in the elevation plane, but significantly increases in the azimuth plane. The switching method can increase the antenna gain slightly, whereas an additional gain reduction occurs in the amplitude tapering method. In short, the uniform array can provide a good scanning range and a higher gain but with the substantially high sidelobes, especially in the elevation plane.

## D. Non-Uniform Conformal Array vs. Planar Array

To analyze the antenna efficiency of the proposed conformal array, a planar array having a projection area identical to the conformal array is used as a benchmark. The planar array also consists of 7×4 elements with the equivalent spacing of $dx$ = 20.8 mm and $dy$ = 25 mm along $x$- and $y$- direction, respectively. Fig. 6 plots the elevation and azimuth scan patterns of this planar array. It is shown that the higher gain above 18 dBi can be achieved over a narrower scanning range of $-66° \leq \varphi_{scan} \leq -30°$, and the worst SLL up to $-3.9$ dB happens at the observation angle of $-58°$ in the elevation scan.

Fig. 6 also reveals that the planar array realizes a slightly narrower scanning range of $-48°\sim48°$ in the azimuth scan but with a smaller scanning loss of 0.56 dB. The maximum directivity of the planar array is 3.78 dB higher than that of the non-uniform conformal array. This is because the normal vectors of all elements in the planar array are in-phase. On the contrary, the normal vectors of the majority of elements in the non-uniform conformal array point toward different directions, making the element itself have a destructive effect on the efficiency of the conformal array. Therefore, to achieve a lower

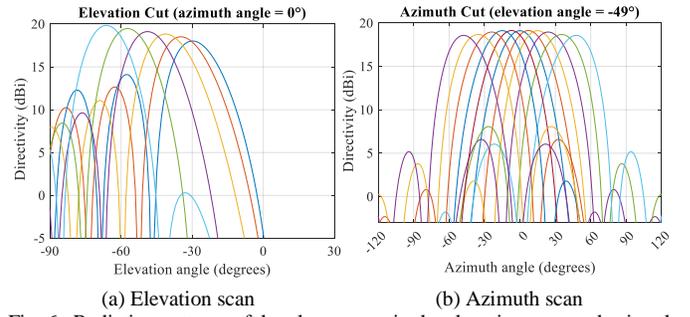

(a) Elevation scan  (b) Azimuth scan

Fig. 6. Radiation patterns of the planar array in the elevation scan and azimuth scan at 5.8 GHz.

TABLE III
OPTIMIZED WEIGHTING COEFFICIENTS OF THE NON-UNIFORM CONFORMAL ARRAY

| No. of elements | Coeff.* | No. of elements | Coeff.* | No. of elements | Coeff.* |
|---|---|---|---|---|---|
| $w_1$ | 0.60004 | $w_{10}$ | 0.60047 | $w_{19}$ | 0.56513 |
| $w_2$ | 0.50536 | $w_{11}$ | 0.65614 | $w_{20}$ | 0.56896 |
| $w_3$ | 0.40776 | $w_{12}$ | 0.35597 | $w_{21}$ | 0.42225 |
| $w_4$ | 0.60416 | $w_{13}$ | 0.37528 | $w_{22}$ | 0.325751 |
| $w_5$ | 0.61405 | $w_{14}$ | 0.54554 | $w_{23}$ | 0.5038 |
| $w_6$ | 0.65088 | $w_{15}$ | 0.55247 | $w_{24}$ | 0.50674 |
| $w_7$ | 0.55968 | $w_{16}$ | 0.473 | $w_{25}$ | 0.30444 |
| $w_8$ | 0.65737 | $w_{17}$ | 0.45924 | $w_{26}$ | 0.30912 |
| $w_9$ | 0.48702 | $w_{18}$ | 0.56513 | $w_{27}$ | 0.44802 |
|  |  |  |  | $w_{28}$ | 0.40781 |

*Noted that the weighting coefficients are non-normalized.

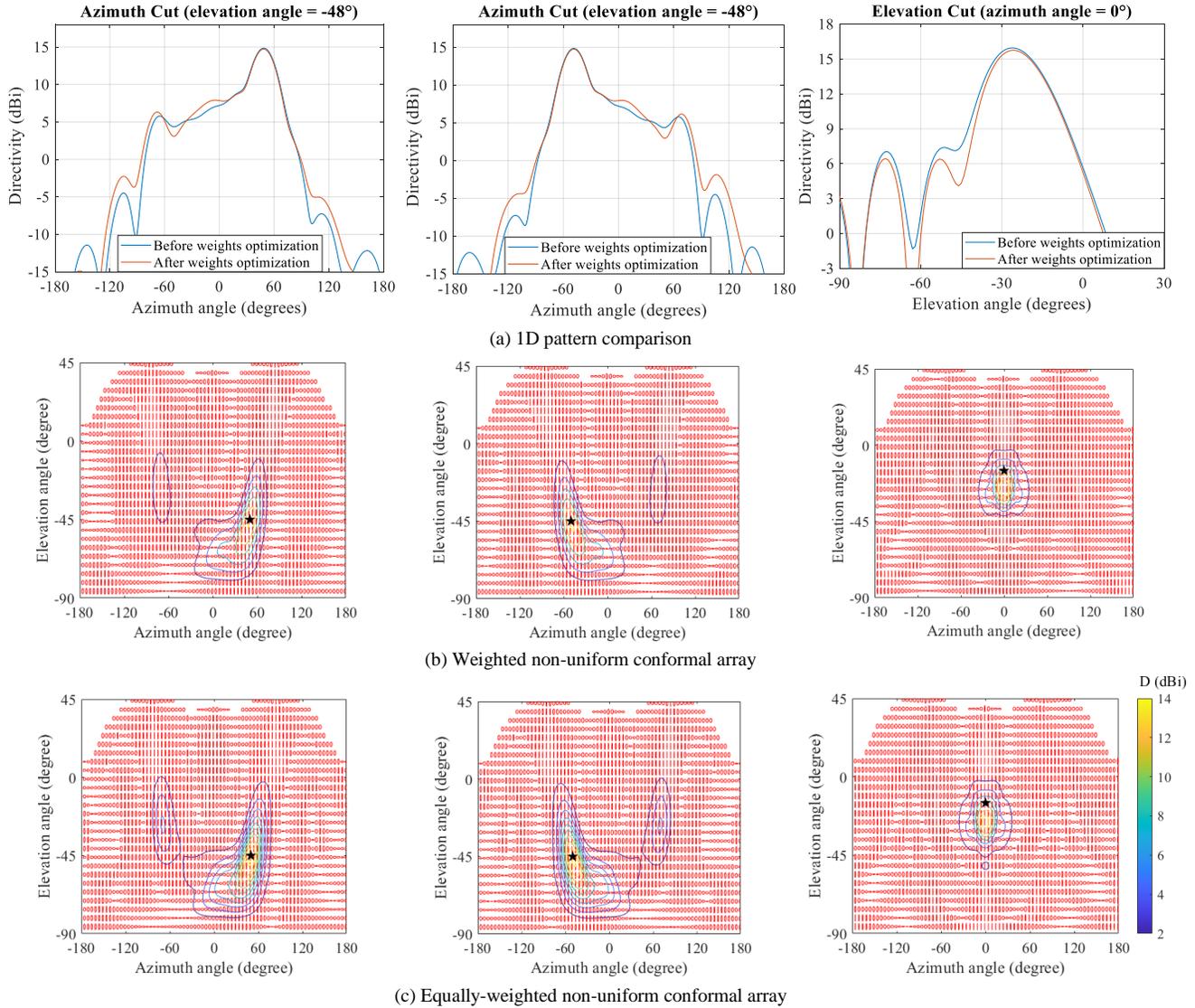

Fig. 7. 1D and 2D farfield patterns of the non-uniform conformal array before and after applying the optimized weighting coefficients to the excitation amplitude in the examples of ($\theta_{scan} = 50°$, $\varphi_{scan} = -48°$) and ($\theta_{scan} = -50°$, $\varphi_{scan} = -48°$) for the azimuth scan, and ($\theta_{scan} = 0°$, $\varphi_{scan} = -27°$) for the elevation scan.

SLL, a wider scanning range, and a better aerodynamic structure, the proposed non-uniform conformal array is a better candidate than the planar array.

*E. Non-Uniform Conformal Array vs. Weighted Non-Uniform Conformal Array*

The non-uniform conformal array discussed in the above Sections II-C and II-D does not take the amplitude weights into account, each element is fed with the same input power for simplicity and fair comparison. In this section, we study the weighted excitation coefficients of the non-uniform conformal array using a convex optimization method described in Section II-A. Table III lists these weighting coefficients after the optimization in order to further improve the pattern shape and the SLL.

Fig. 7 compares the farfield patterns of the non-uniform conformal array before and after applying the optimized weighting coefficients to the excitation amplitude in the examples of ($\theta_{scan} = 0°$, $\varphi_{scan} = -27°$) for the elevation scan, ($\theta_{scan} = 50°$, $\varphi_{scan} = -48°$) and ($\theta_{scan} = -50°$, $\varphi_{scan} = -48°$) for the azimuth scan, respectively. As shown, within an identical 2D scanning coverage, the SLL is more suppressed in the elevation scan and further moves outside the desired azimuth scanning range by employing the weighted excitation to the array. Although the antenna gain is slightly decreased, the large sidelobe tail is obviously shrunk and the pattern is well re-shaped after using the optimization algorithm. Therefore, it is concluded that the optimized weighting coefficients indeed can bring some benefits to the non-uniform conformal array, but with a sacrifice of system simplicity and implementation cost. In this regard, we eventually choose to implement the equally-weighted non-uniform conformal array as a proof of concept, as discussed in the following section.

## III. IMPLEMENTATION OF NON-UNIFORM CONFORMAL ANTENNA ARRAY

In this section, a non-uniform conformal antenna array

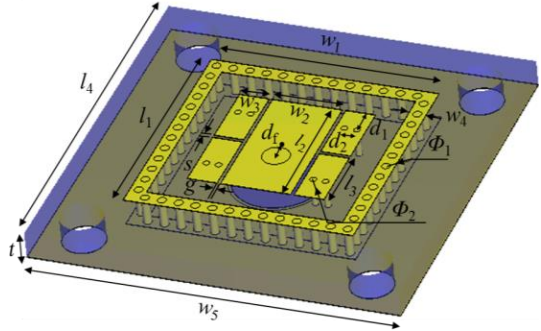

Fig. 8. Configuration of the modified miniature microstrip antenna element.

TABLE IV
PARAMETERS OF THE OPTIMIZED MINIATURE MICROSTRIP ANTENNA

| Par. | Value (mm) | Par. | Value (mm) | Par. | Value (mm) |
|---|---|---|---|---|---|
| $w_1$ | 10.8 | $d_1$ | 1.26 | $l_2$ | 6.44 |
| $w_2$ | 3.3 | $d_2$ | 0.59 | $l_3$ | 3.12 |
| $w_3$ | 1.38 | $d_f$ | 0.7 | $\Phi_1$ | 0.4 |
| $w_4$ | 0.8 | $s$ | 0.19 | $\Phi_2$ | 0.3 |
| $l_1$ | 10.8 | $g$ | 0.19 | $t$ | 1.27 |
| $l_4$ | 17.5 | $w_5$ | 17.5 | | |

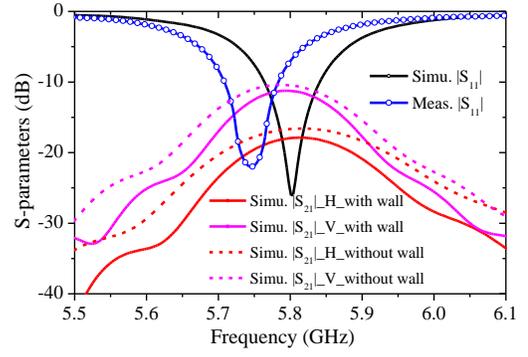

(a) Input impedance matching and mutual coupling

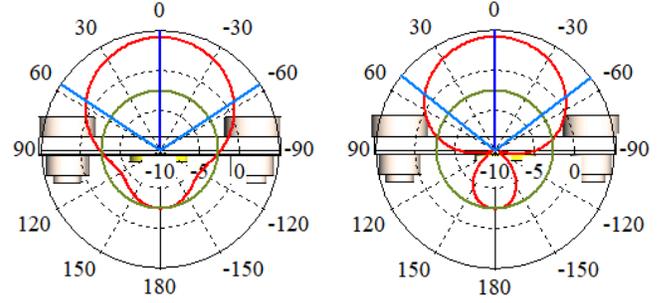

(b) H-plane pattern  (c) E-plane pattern

Fig. 9. Simulated S-parameters and radiation patterns (at 5.8 GHz) of the modified miniature microstrip antenna element.

operating in ISM band, as proposed in Section II-B, is implemented and modelled with the aid of a full-wave EM simulation tool *CST Studio Suite*, including the corresponding BFN board design to control the amplitude and phase of each element. First, a compact single-layer patch antenna with a specified directive radiation is utilized as the array element to achieve the critically small element spacing and reduce the mutual coupling. Then, the whole onboard array system is designed and the array performance is examined when the proposed array is conformed on the bottom front of the UAV.

*A. Array Element*

Aside from the standard PCB fabrication tolerance, the antenna element design is challenging because the antenna size should be as small as possible to meet the minimum spacing of ~18 mm (only $0.35\lambda_0$), the beamwidth has to be wide in order to achieve the wide beam scanning, and the element gain should be as high as possible to guarantee a good radiation efficiency. Therefore, there is a trade-off between the far-field radiation, isolation, compact size, and simple structure for the array element design.

According to our study, the miniature single-layer patch antenna in [27] can be modified for our purpose. As illustrated in Fig. 8, in addition to the patch size, the slots and the metallic pins enable the increase in the antenna design's degree of freedom and a significant reduction in the resonant frequency of the original patch. The outer metallic wall and periodic posts surrounding the center patches are used to suppress the surface wave and reduce the mutual coupling to its neighbors, while keeping the patch size unchanged. The substrate used here is TMM10i with a dielectric constant of 9.8 and a tangential loss of 0.002 at 10 GHz. The designed parameters are specified in Table IV. It is observed that the overall patch size is merely 6.44×6.44 mm² ($0.12\lambda_0 \times 0.12\lambda_0$) for the resonant working frequency of 5.8 GHz.

Fig. 9 depicts the simulated farfield patterns of the proposed array element at 5.8 GHz and compares its S-parameters with those of the reference antenna without the outer conducting wall. Two elements are placed with a spacing equal to 18 mm ($0.348\lambda_0$) to examine their isolation. As can be seen, regardless of the horizontal and vertical arrangements of the two elements, the proposed antenna is still resonant at 5.8 GHz with an input impedance bandwidth of 77 MHz and 1~2 dB coupling reduction between two elements, when compared to the reference antenna. The boresight gain of the designed element is 4.11 dBi at 5.8 GHz with a 1dB gain flatness bandwidth from 5.75 GHz to 5.86 GHz and a low front-to-back ratio of 6.73. The antenna beamwidth is 112.6° and 104.6° in H- and E-planes, correspondingly. It should be highlighted that the pattern shape of the proposed antenna element is well matched to the $\cos^2$ pattern in Section II, making a straightforward comparison between the fast synthesis and the full-wave simulation results.

*B. Conformal Antenna Array*

Based on the optimized array topology, a non-uniform conformal array consisting of 28 elements is implemented with the array elements designed in Section III-A. As illustrated in Fig. 10(b), the element position and normal vector defined from the theory in Section II-B can be easily realized in the implemented design. The slight difference between the fitting surface and the actual UAV surface can be compensated by the designed dielectric support mask, which is able to conform the flat ground plane of all array elements and the UAV surface on its two opposite sides. To consider the aerodynamic and mechanical performance of the entire array on the UAV, the support mask should be as thin as possible, and $dy_1$ is modified to 18 mm so that the protruding UAV bottom remains compact.

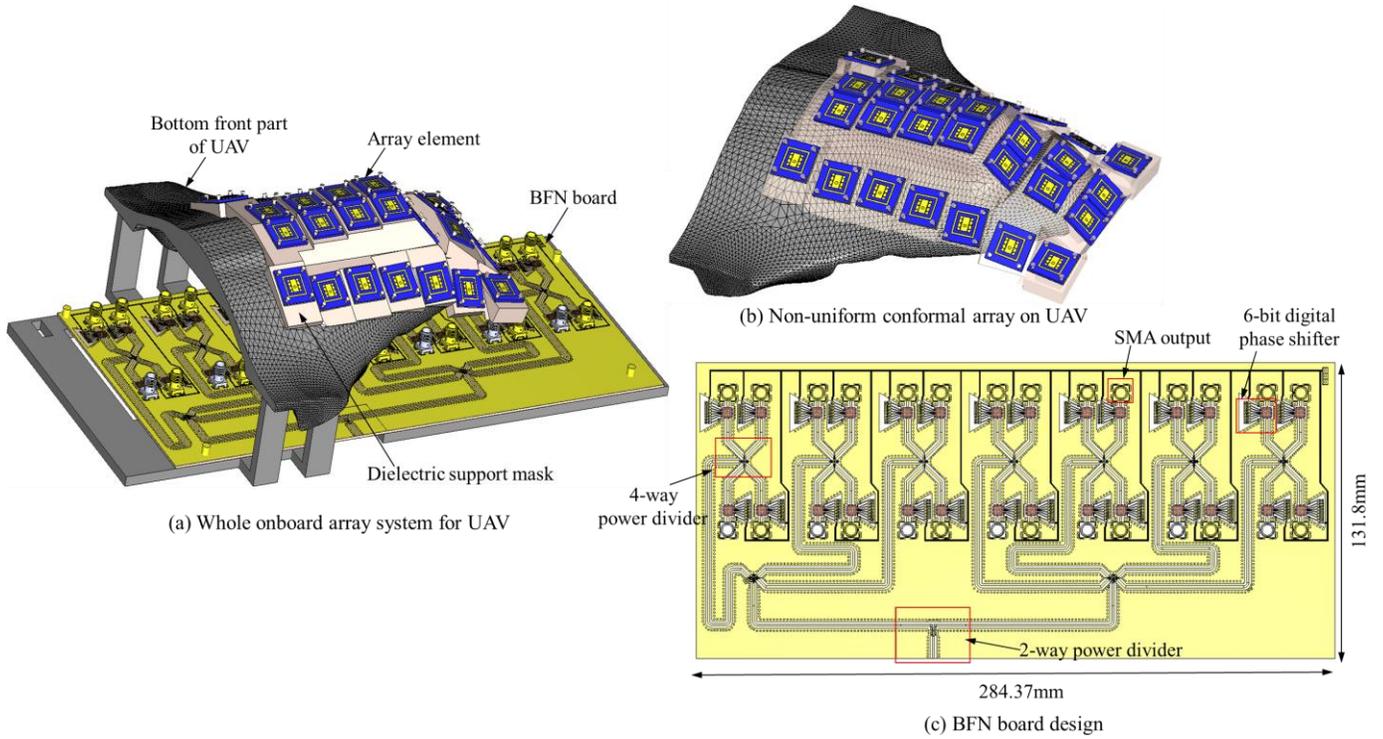

Fig. 10. Configuration of the proposed non-uniform conformal array on UAV, the corresponding BFN board, and the whole onboard array system.

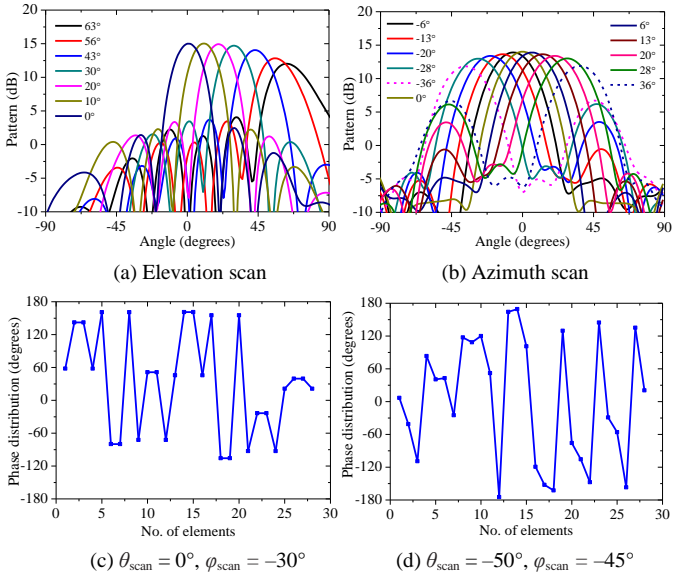

Fig. 11. Simulated beamscanning performance of the implemented non-uniform conformal array in elevation and azimuth, and the theoretical phase distribution for the steering angle of ($\theta_{scan} = 0°$, $\varphi_{scan} = -30°$) and ($\theta_{scan} = -50°$, $\varphi_{scan} = -45°$), respectively.

The simulated beamscanning performance of the implemented non-uniform conformal array is plotted in Figs. 11(a-b) and compared with the theoretical patterns in Figs. 4(d) and (f). The phase distribution of the array elements required to achieve the specific steering angle can be calculated using the expression $e^{-jk r_i \cdot r_s}$. Here, the ideal phase distribution for the critical steering angles ($\theta_{scan} = 0°$, $\varphi_{scan} = -30°$) and ($\theta_{scan} = -50°$, $\varphi_{scan} = -45°$) is shown in Figs. 11(c-d). By applying the calculated phase distribution to 28 elements, a wide 2D beam coverage of >56° in elevation scan and −28°~28° in azimuth scan can be achieved within a small scan loss of 2.2 dB and 1.1 dB, respectively. The narrowed azimuth scanning range mainly results from the stronger mutual coupling along that direction which is not considered in theory, and this phenomenon is consistent with the higher $|S_{21}|\_V$ in Fig. 9(a). However, if the phase distribution for a wider steering angle is applied to the full-wave model, a wider azimuth scan is still achieved but with a larger scan loss, as shown with the dash lines in Fig. 11(b). Overall, regardless of the slight increase in SLL (grating-lobe effect due to the enlarged $dy_2$), the scanning pattern shape of the implemented array is consistent with the theoretical case. In particular, these high sidelobes in azimuth still remain outside the main scanning range. Hence, the proof-of-concept design validates the feasibility and accuracy of the proposed array synthesis method with the in-house algorithm and the trivial effects from the final geometry modification.

*C. Onboard Array System*

The entire architecture of the onboard array system using the beamforming technique for UAV application is demonstrated in Fig. 10(a). The architecture consists of a conformal layer for the antenna array, a dielectric support mask coated on the bottom front of the UAV, and a planar layer for the BFN board to realize the required amplitude and phase for each element.

Specifically, the BFN board comprises 28 digital phase shifters, nine 4-way power dividers and one 2-way power divider that allocate the equal phase and amplitude of the RF signal injected to the corresponding phase shifter. Since the digital phase shifters are active, an extra DC power supply is needed to realize the bit control. To save space on the layout, DC signals are also designed on the same PCB panel as the RF signals. It is verified in simulations that the mutual coupling between the RF transmission lines and the DC lines is negligible by the use of grounded co-planar waveguide (CPWG) transmission lines. Moreover, to minimize the phase difference among the transmission ports, the same physical length of RF transmission line segments between the phase shifters and power dividers are used. The size of the BFN board is roughly twice of the antenna array, a higher antenna gain is thus envisaged in the later measurement.

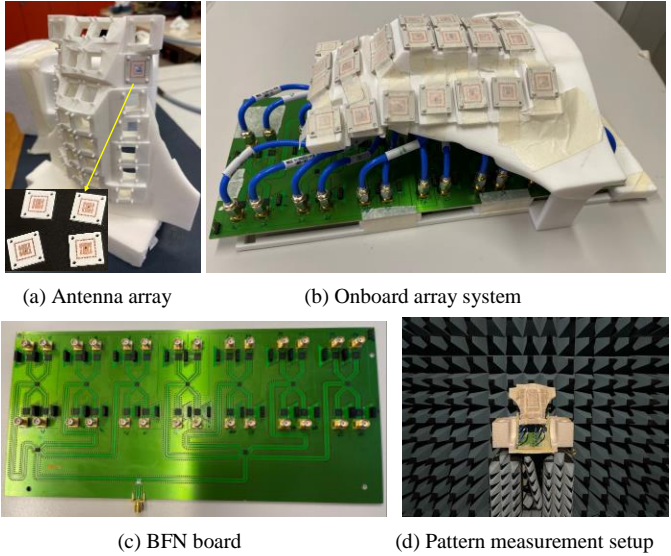

(a) Antenna array  (b) Onboard array system

(c) BFN board  (d) Pattern measurement setup

Fig. 12. Fabricated prototypes of the proposed non-uniform conformal array, BFN board, and the entire onboard system and measurement setup.

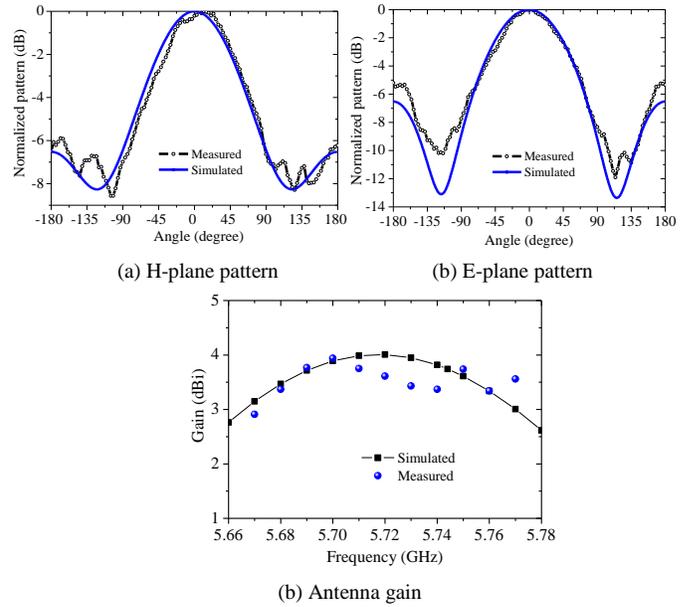

(a) H-plane pattern  (b) E-plane pattern

(b) Antenna gain

Fig. 13. Measured radiation patterns of the fabricated array element at the operating frequency of 5.72 GHz and antenna gain over frequency.

## IV. MEASUREMENT AND RESULT DISCUSSION

### A. Fabricated Prototypes

Figures 12(a-c) present the prototypes of the proposed non-uniform conformal array, the BFN board, and the integrated onboard system fabricated in collaboration with ACI (Atelier de fabrication de Circuits Imprimés) workshop and AFA (Atelier de Fabrication Additive, 3D printing) workshop in our university. The dielectric support mask and the UAV model are 3D printed using SLS (laser powder fusion) technology for demonstration. The farfield pattern measurement is performed in the anechoic chamber, as shown in Fig. 12(d).

In order to achieve a 360° dynamic phase range at the specified frequency, a digital phase shifter with a small step increment, low phase error, and low insertion loss is required. HMC1133LP5E is a GaAs MMIC 6-bit digital phase shifter that is ideal for an arbitrary beamforming in the radar and satellite applications, operating from 4.8 GHz to 6.2 GHz with a 5.6° digital step size. Hence, it can be utilized to implement the BFN board. Concerning the 4-way power divider, an MMIC surface mount power splitter WP4A+ is chosen since it features a low insertion loss, a low phase unbalance, a low amplitude unbalance, and a good VSWR at the specified frequency range. Similarly, an ultra-small ceramic power splitter SCN-2-65/65+ is selected for the 2-way power divider because of its extremely small amplitude unbalance, low phase unbalance, good insertion loss and return loss.

### B. Performance of Antenna Element

The measured input impedance matching of the fabricated array element is plotted in Fig. 9(a) and compared with the simulated result. The resonant frequency shift is mainly caused by the fabrication tolerance on the slots and patch size. These parameters are quite sensitive since the 80 MHz antenna bandwidth is narrow. Therefore, the pattern measurement will be conducted for the actual operating frequency at 5.72 GHz. The measured radiation patterns of the array element in H-plane and E-plane are presented in Fig. 13, and they are in good agreement with the simulated patterns, showing a wide beamwidth of 110° and 101° in the corresponding plane as well. The measured 1dB gain reduction bandwidth is more than 100 MHz with a maximum of 3.94 dBi, and is consistent with the simulated gain.

### C. Performance of Active Onboard Array System

Figure 14 plots the measured farfield patterns for the elevation and azimuth scan at 5.72 GHz, respectively. It reveals that the proposed non-uniform conformal array associated with its compact onboard system can achieve a scanning range of 0°~51° for elevation and -28°~28° for azimuth, within a very small scanning loss of 2.2 dB and 0.5 dB, respectively. The measured maximum antenna gain is around 21.8 dBi, which is significantly higher than the result in Fig. 11(a) by 6.7 dB. This is because the BFN board is not taken into account in the simulation, and this large flat panel serves as an additional reflector to improve the antenna gain. Also, the presence of the close BFN board results in a little higher sidelobes and pattern difference in the measurement. Nevertheless, the SLL in the elevation scan is comparable to the theoretical calculation, and the sidelobes in the azimuth scan are still out of the targeted scanning range, thus do not affect the mainbeam performance in that plane. The slight difference in beam direction might be caused by the actual mounting position of the array elements on the UAV and the misalignment between the AUT (antenna under test) and the reference horn antenna. In conclusion, the proposed proof-of-concept design with the use of the fast and efficient array synthesis method has been experimentally validated.

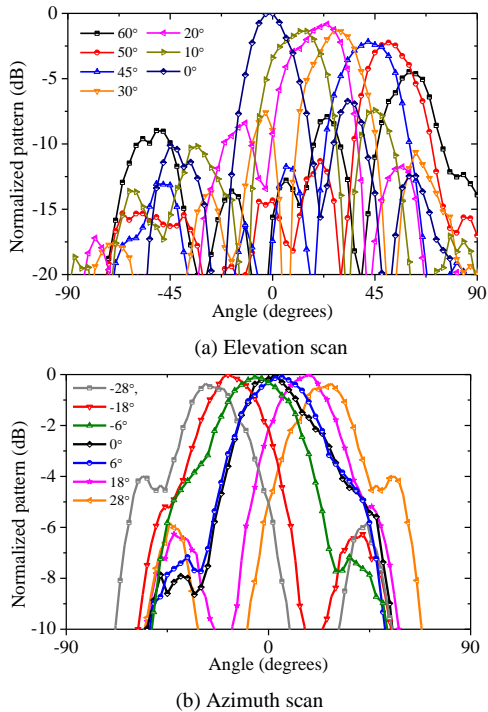

(a) Elevation scan

(b) Azimuth scan

Fig. 14. Measured farfield patterns of the fabricated onboard array system for the elevation and azimuth scan at 5.72 GHz.

## V. CONCLUSION

With the help of an improved algorithm for the radiation calculation of a non-canonical conformal array, a miniaturized non-uniform conformal array for the sophisticated UAV surface is synthesized in ISM band, and its array geometry can be optimized with a reduced computation cost. The non-uniform array consists of only 7×4 discrete elements which are the modified single-layer microstrip antennas with an extremely small dimension of $0.12\lambda_0 \times 0.12\lambda_0 \times 0.025\lambda_0$ and improved isolation among the elements. The proposed non-uniform array, in principle, can achieve the scanning range of >50° and ±49° in the elevation and azimuth plane, respectively, with the maximum directivity of 15.9 dBi and improved SLLs. To prove the concept and the design method, an active beamforming onboard system with the proposed non-uniform conformal array is designed and implemented. The designed array element can achieve the specified $\cos^2$ pattern with a linear polarization. The BFN board is elaborately designed on a single PCB layer to realize a compact system. Also, the antenna array can be easily conformed on the UAV body by the use of a thin 3D-printed support mask. The measurement results show a good consistency with the full-wave simulation results, with an extra gain due to the presence of the close BFN board.